# Observation of the formation of long-range order during superradiant emission in a high-density e-h system

Peter P. Vasil'ev[1,2] & Ian H. White[1]

[1] *Centre for Photonic Systems, Department of Electrical Engineering, University of Cambridge, 9 JJ Thomson Avenue, CB3 0FA, UK*

pv261@cam.ac.uk

[2] *PN Lebedev Physical Institute, 53 Leninsky Prospect, Moscow 119991, Russia*

The transition of atomic, electronic, or e-h systems to quantum (condensed, superfluid or superconducting) phases is an extraordinary process when both microscopic and macroscopic properties of the system undergo fundamental changes. The investigation and understanding of such processes has been carried out for over a hundred years. Bose-Einstein condensation (BEC), i.e. accumulation of a microscopic number of particles at the ground energy state, has been observed in a number of physical systems, including super-cold atomic gases and solid-state quasiparticles [1-5]. In some cases the quantum states of matter are the results of an equilibrium condensation. The most typical example is the superconducting state of Cooper pairs in metals. However, in many cases the lifetime of particles or quasiparticles under study is finite, ranging from picoseconds to milliseconds. And even in this case condensation is possible, provided that the lifetime of the particles is much longer than the characteristic time which they require to scatter with each other. It is generally understood at present that coherence and off-diagonal long-range order in the density matrix are the most important characteristic features of condensed quantum states [6,7]. Phase coherence established over microscopic distances is an intrinsic property of all Bose condensates. Experiments featuring interference between either two condensates or two separate parts of the same condensate are often considered as more convincing than observation of the macroscopic occupation of the ground state [6]. The formation of long-range order and spatial phase coherence can be measured by studying interference patterns using either a Michelson interferometer or Young's double slit arrangement. Here we present experimental results of observation of long-range temporal and spatial coherence of a macroscopically ordered state of a high-density e-h system in GaAs during the generation of femtosecond superradiant (SR) emission at room temperature.

Dicke superradiance[8] (collective spontaneous recombination) exhibits many characteristics of a quantum phase transition [9,10]. One of the fundamental features of the SR is the mutual phasing of emitters without which the generation of the SR is not practically possible[11]. This self-organisation originates from an exchange of photons of the internal electromagnetic field. The most striking consequence of interactions among particles is the appearance of new phases of matter whose collective behaviour bears little resemblance to that of a few particles. Developing of the collective superradiant state in the medium implies the build-up of a macroscopic polarization, which in turn suggests the formation of an ordered state. After intensive theoretical studies[12] SR emission has been experimentally observed in many types of media - gases, solids, and semiconductors, including quantum dots and excitonic condensates[11,13].

The generation of SR emission in semiconductors has been achieved in the form of high-power femtosecond pulses from bulk GaAs/AlGaAs laser-type structures in late 1990s [14,15]. The ability of those structures generate conventional laser emission along with SR pulses raised the issue of observed femtosecond emission to be just an extreme case of normal lasing. However, further experimental and theoretical studies[16-18] have explicitly demonstrated that the characteristic features of SR emission are completely different from those of lasing. The fundamental difference originates from the distinction of intrinsic properties of the semiconductor medium under lasing and SR generation. The optical emission from a semiconductor laser is coherent, whereas the semiconductor medium remains incoherent. By contrast to lasing, both electromagnetic field and electron-hole system are coherent during the SR generation. In this paper we experimentally demonstrate using the Michelson and Young's double slit experiments that e-h system exhibits long range spatial coherence both in the longitudinal and lateral directions of the semiconductor structure active region. The present results are consistent with the concept of the formation of coherent BCS-like e-h state during the SR generation in a semiconductor medium, which has been developed previously[11, 16-18].

**Experimental procedure**

Two sets of bulk GaAs/AlGaAs heterostructures with different configurations were studied. The structures of both sets have a 3-section geometry, with the two end sections being the amplifying medium with optical gain and the centre one being an electrically controllable saturable absorber. Figure 1 illustrates the device under test. The *i*-GaAs active layer sandwiched by *p*- and *n*-AlGaAs layers is typically 0.2 microns thick. Nanosecond current pulses applied on the gain sections create e-h concentrations in the range of $10^{17} - 10^{19}$ cm$^{-3}$, depending on the current amplitude. The application of the variable d.c. reverse bias on the centre section allows for a precise control of the optical absorption of photons travelling in the active layer. The cleaved facets of the structure have the power reflection coefficient of around 0.32 and provide an optical feedback. Spontaneous emission and normal lasing can be generated from the device at different values of *I* and *V*. For instance, the typical e-h threshold concentration for lasing with *V*= 0 is around (1.0-1.5) x $10^{18}$ cm$^{-3}$.

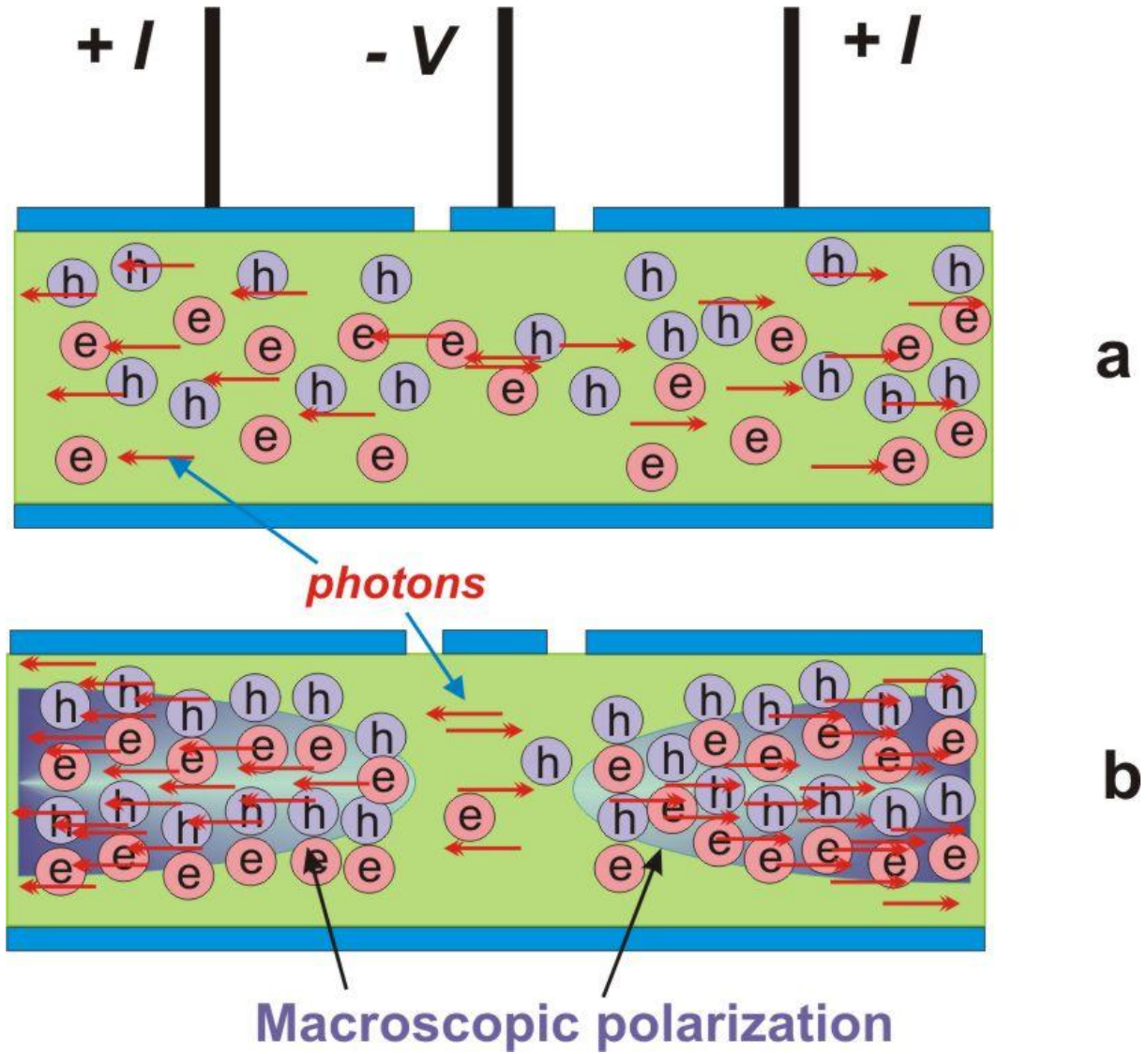

Figure 1. A sketch of the 3-section semiconductor device. **a**, lasing or spontaneous emission. **b**, superradiance. The end sections are driven with the forward pulsed current *I*, while the reverse d.c. bias – *V* is applied on the centre section. The carrier density for **b** is 3-6 times larger than for **a**. The two regions of macroscopic polarization with ordered individual e-h dipoles are formed at the ends of the device during the SR emission generation.

Due to different processes of ultrafast relaxation in the semiconductor medium (e-e, e-h, h-h collisions and interactions with phonons), electron-hole pairs remain incoherent with each other. Typical values of the polarization relaxation time are $10^{-14}$ - $10^{-13}$ s at room temperature. As a result, there is no any order in the e-h system (see Fig 1**a**).

Increasing the bias *V* increases the absorption and can prevent onset of lasing. However, for large enough values of *I* and *V*, the SR state can develop in the active layer (Fig. 1**b**). The conditions and mechanism of the formation of this state have been investigated previously[11, 16-18]. Typical e-h densities in this case can exceed 6 x $10^{18}$ $cm^{-3}$. As suggested elsewhere[16] an ensemble of the e–h pairs with zero total wavevector (the cooperative state) emerges during the initial phase of the build-up of a SR pulse. The photons of the internal electromagnetic field then establish correlations within the e-h system. The longest wavelength photons build correlations and coherency in the ensemble due to the absorber section and special profile of the optical gain in the device[17]. The macroscopic polarization implies ordering of individual dipoles of e-h pairs. The cooperative state is highly spatially non-uniform along the longitudinal axis because of the large optical gain and the resulting exponential increase in the electromagnetic field towards the sample ends. The state therefore locates itself near the chip facets as illustrated in Fig. 1**b**. Its coherence can be easily probed by monitoring the output emission from the facet using a Michelson interferometer. We performed such experiment with the first set of the devices. These have straight 5 microns wide active layers with the total length of 100 microns. The absorber section is 10 μm, whereas the gain sections are 30 μm long. The results of the experiment are presented in the next Section.

The second type of the experiments deals with measurement of the spatial coherence of the ordered state in the lateral direction. We use the famous Young's double-slit experiment[19], which has recently become a benchmark demonstration of macroscopic long-range spatial coherence and off-diagonal long range order of condensed quantum states[20,21]. The second set of the semiconductor structures has a tapered geometry with the active layer width increasing from 5 to 30-40 μm depending on the sample. The schematic of the experimental set-up is shown in Fig. 2. In fact, we observe with a CCD video camera interference patterns of the spontaneous, lasing, and SR emissions from the semiconductor structure passing through the double slit sample. Since the intensity of spontaneous, lasing, and SR emissions varies enormously, the variable optical attenuator was used to ensure the linear detection of the camera. Three sets of double slits of different sizes (5, 10, and 15 μm) with the separation *r* increasing from 5 to 200 μm were used. The spatial coherence and long range order can be measured by studying of the interference pattern of the emission originating from two different locations of the emitting aperture of the device. The visibility of this interference pattern measures the phase coherence between the two locations. The evolution of the phase coherence inside the semiconductor medium at different dynamic regimes (spontaneous, lasing, or SR) can be mapped out by changing the slit separation *r* and determination of the first order correlation function $g^{(1)}(r)$.

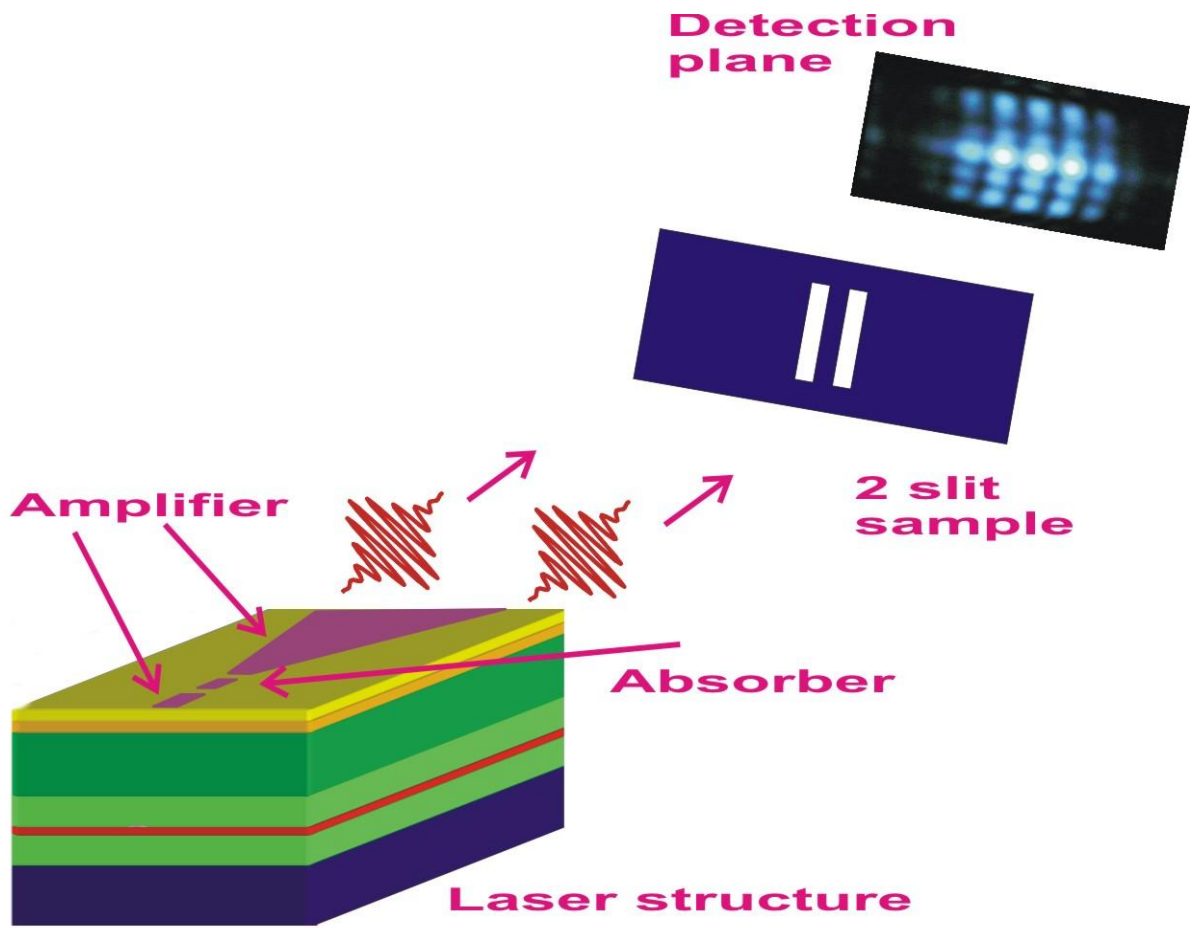

Figure 2. Schematic of the Young's double slit experiment with a 3-section tapered device. Two photons emitted from two different locations of the output aperture interfere at the double slit sample. The interference pattern is detected by the CCD camera and is further processed and analysed by a computer.

**Long range order coherence along the longitudinal axis**

The first set of measurements was carried out with a Michelson interferometer. We used the same interferometer which had been employed for SHG autocorrelation measurements of pulsewidths of ultrashort optical pulses[22]. Scanning of the arm of the interferometer with a micron and sub-micron accuracy results in fringe-resolved autocorrelation traces which allows for precise monitoring of phase coherence of the optical emission under test. It is well-known that a fringe-resolved autocorrelation trace of any laser emission exhibits a *single* peak at zero delay. Its width determines the coherence time of laser emission and is inversely proportional to the spectral bandwidth. For mode-locked pulses there exist additional identical peaks separated by the round trip time of the laser cavity[22]. SHG autocorrelation traces of SR emission from short length (100 μm) devices look very different. Figure 3 presents typical autocorrelation traces. They are unique for any emission from semiconductor media.

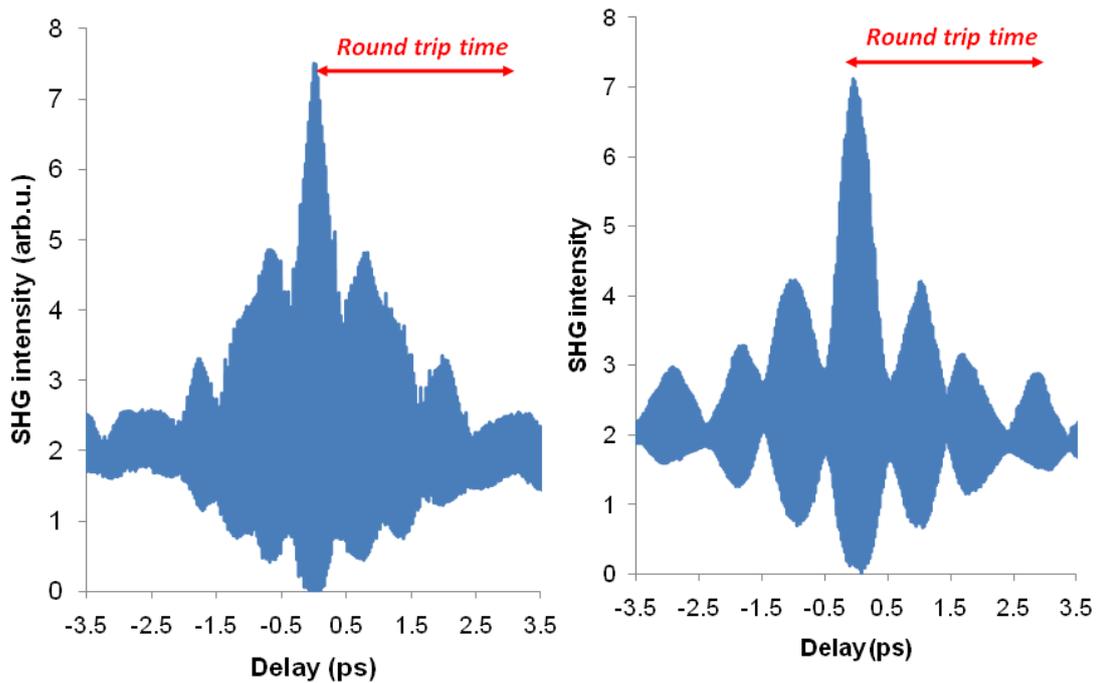

Figure 3. Michelson interferometer traces illustrating coherence in longitudinal direction. The device length is 100 microns which corresponds to the round trip time of 3.1 ps. The characteristic time $T_2$ is in a range of 20-100 fs. SR emission travels along the longitudinal axis between the chip facets reabsorbing and reemitting again. Multiple peaks and fringes at nonzero delays confirm the coherent interaction of the electromagnetic field with the semiconductor medium.

The characteristic features of the traces, which distinguish them significantly from these of lasing, are multiple peaks and presence of fringes at nonzero delays at times much longer than $T_2$. The single peak at zero delay of a correlation trace of a laser emission is a result of incoherent amplification of the emission by the active medium of the laser. Indeed, e-h pairs remain coherent with the travelling resonant electromagnetic field within a time determined by the transverse (polarization) relaxation time $T_2$. Numerous ultrafast relaxation processes within the semiconductor result in a quick quenching of the coherence and the e-h system 'forgets' the phase information of the electromagnetic field. The amplifying medium in any laser remains incoherent while the electromagnetic field is coherent. It should be pointed out here that a coherent medium within ~ $T_2$ times can be created by an external resonant pumping. In this particular case, some interesting phenomena like photon echo, self-induced transparency, π- and 2π-pulse generation, etc. can be observed[23].

The SR correlation traces in Fig. 3 can be readily understood if one takes into account the existence of two regions of the ordered e-h state (macroscopic polarization) as illustrated in Figure 1. Correlations in the e-h ensemble induced at the early stages of the development of the superradiant state, photon mediated collective pairing and e-h condensation towards the band gap energy lead to the formation of the coherent BCS-like state as described elsewhere[16-18]. The radiation once emitted from a certain region of this state can be absorbed and then reemitted by another part of it. Interference fringes which are present in Fig. 3 all the way through the round trip time suggest the retention of the long-range spatial coherence in the e-h system. The observed multiple peaks in Fig. 3 imply an oscillatory process of the energy exchange between the electromagnetic field and the resonant (e-h) medium similarly to Bloch oscillations in a two-level system.

**Spatial coherence and the first order correlation function**

For the second set of the experiments we used a number of spatially multimode tapered structures. Before performing the Young's double slit experiment we studied the dynamics of lasing and SR generation from different regions of the semiconductor structures. The near-field emission dynamics was investigated using a single-shot streak camera with about 1.5 ps temporal resolution[11]. It is well-known that strong optical nonlinearities of the semiconductor media result in the appearance of different types of nonlinear optical phenomena, including self-focusing, optical phase conjugation and Raman scattering. Lateral instability of the optical field in the active region, especially in broad-area lasers (wider than 2-6 $\mu$m depending on the wavelength), leads to the development of the filamentation of the near-field emission pattern. This is a result of the independence of individual spatial modes and spatial incoherency of the active medium. The near-field inhomogeneity, instabilities and filamentation of the laser emission were always observed in our experiments as shown in Figure 4.

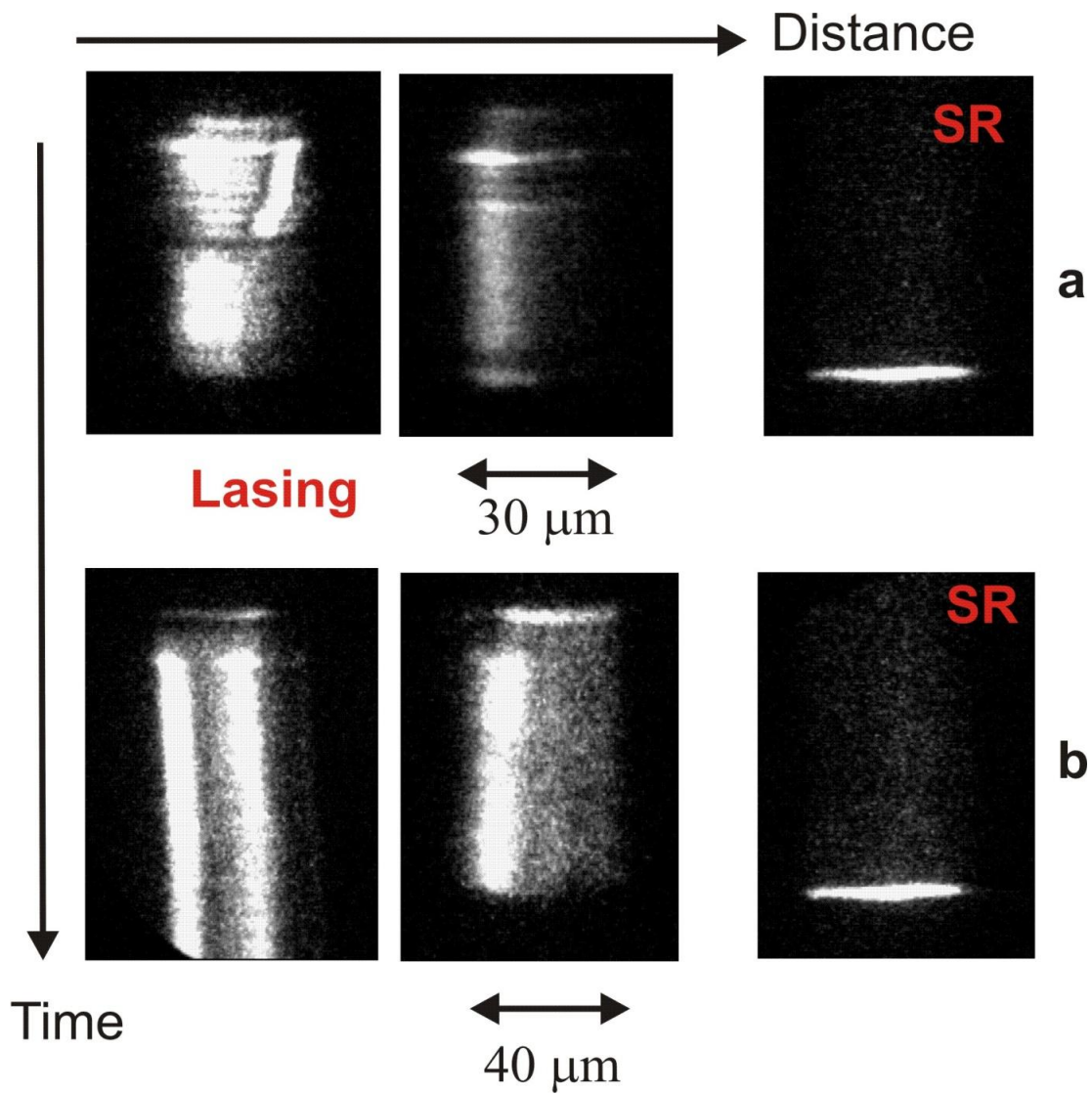

Figure 4. Near-field emission dynamics in lasing and SR regimes. The output aperture of the tapered structures is 30 (**a**) and 40 (**b**) microns. The total time span on all pictures is around 3 ns. The images of the SR pulses do not represent the actual pulse duration which lies in the sub-picosecond range.

It is clearly seen that the laser intensity varies strongly both in space and time. By contrast, SR emission is always generated simultaneously from the whole aperture of the device. Let us now show that the SR state exhibit an ultimate spatial coherence exceeding that of the laser emission from the same tapered structures.

Figure 5 presents two sets of raw pictures obtained using the Young's double slit arrangement shown in Figure 2. It is quite obvious that the visibility of the fringes for the SR emission is larger than that for lasing. The difference in the visibility reflects the difference in the phase coherence.

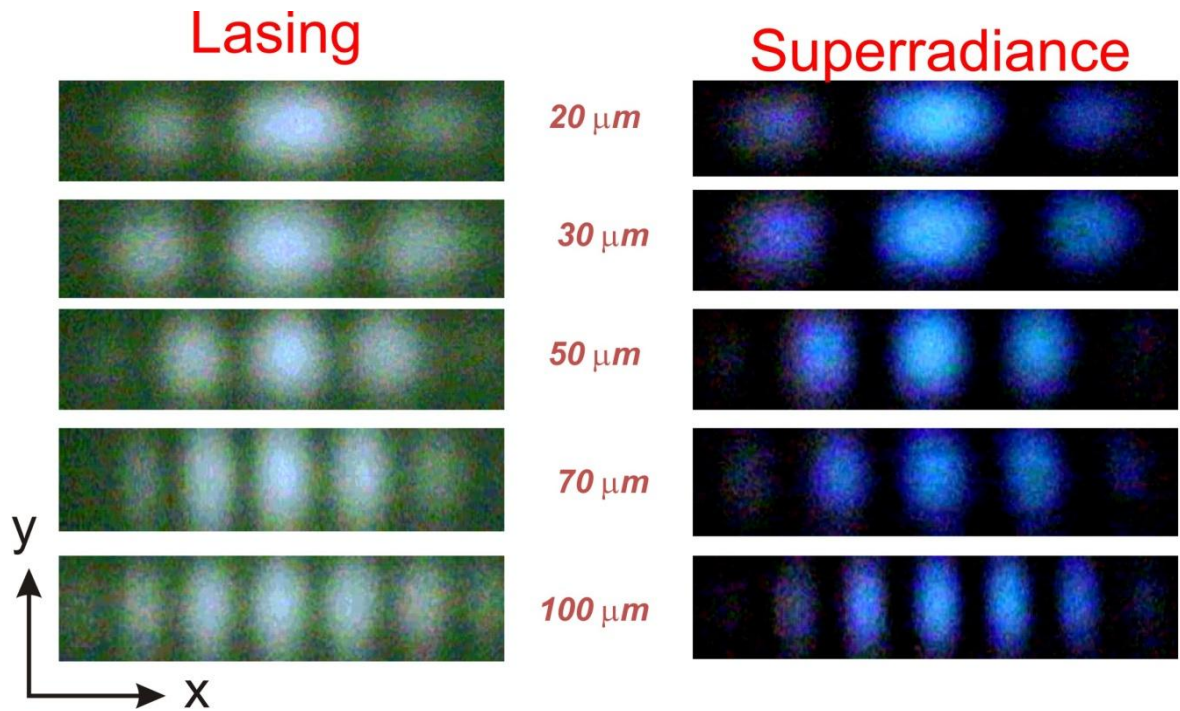

Figure 5. Raw images of 2-slit interference patterns for lasing and SR at different slit separations. The wavelength of the lasing and SR emission is 876 and 887 nm, respectively. The interference patterns of the spontaneous emission are virtually unobservable or nonexisting.

The first-order spatial correlation function $g^{(1)}(r)$ quantifies the off-diagonal long-range order[3,7,20]. In addition, $g^{(1)}(r)$ is considered as a measure for the degree of condensation in momentum space[20]. The correlation function $g^{(1)}(r)$ can be determined by analyzing the interference patterns of Fig. 5. In general, the interference of light through the two slits consists of a cosine oscillation imposed on a *sinc* function envelope. To obtain $g^{(1)}(r)$ in our case, we integrate the intensity of the CCD image $I(x,y)$ over a narrow strip along the y axis and fit it with the theoretical dependence along the x axis[24]

$$I(x) = I_1(x) + I_2(x) + 2g^{(1)}(r)\sqrt{I_1 I_2}\cos(\varphi(x) + \varphi_{12}), \quad (1)$$

where

$$I_{1,2} = |E_{1,2}|^2 sinc^2\left(\frac{x-x_0\pm d/2}{X}\right), \varphi(x) = \frac{2(x-x_0)}{X_c}, X = \frac{2D}{k\delta}, X_c = \frac{2D}{kd} \quad (2)$$

The description of the geometric parameters $d$, $D$, $x_0$, $\delta$, as well as the parameters $I_1$, $I_2$, $\phi(x)$, $\phi_{12}$ and the fitting procedure can be found elsewhere[21]. Figure 6 presents the experimental data and the approximations $I(x)$ using Eqs. (1)-(2).

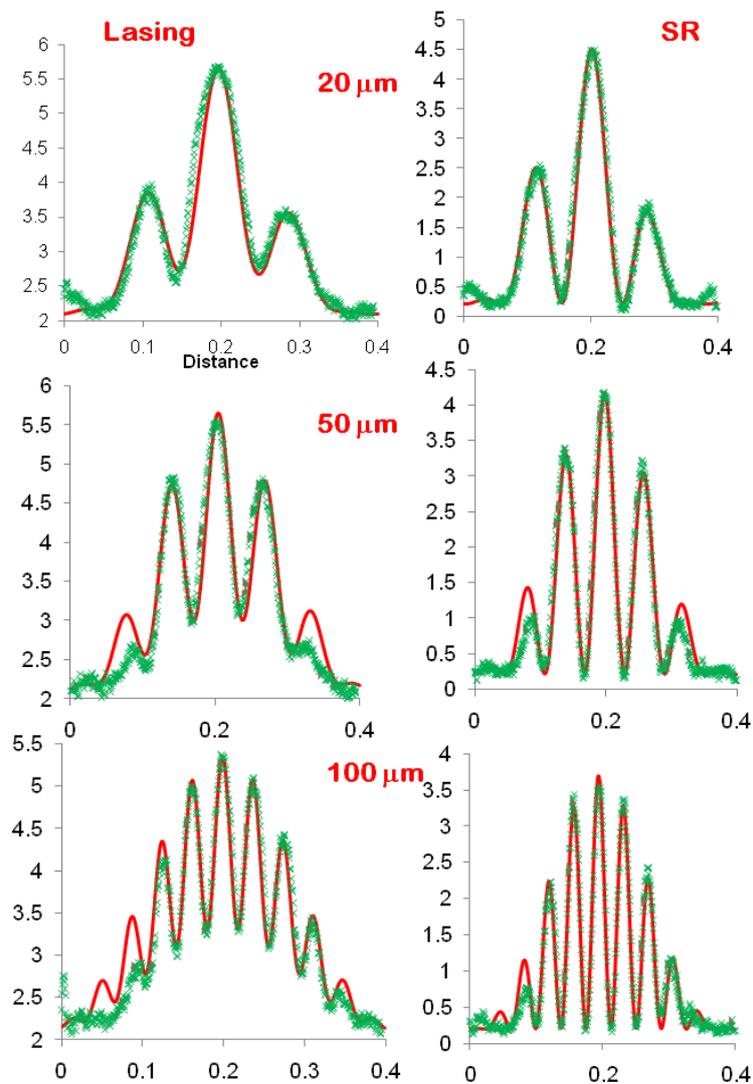

Figure 6. Measured intensities I(x) (green crosses) and fittings (red lines) by Eq.(1)-(2) for lasing and SR. The slit separations were 20, 50, and 100 microns.

Our calculations have shown that the experimental data can be fit very well by the theoretical curves. The only obvious discrepancy occurred at the wings of *I(x)* for the laser emission. This can be caused by strong inhomogeneities of near- and far-field distributions of the laser emission, whereas Eqs. (1)-(2) is relevant for an ideal case. By contrast, the fitting of *I(x)* for SR was always more accurate (see Fig. 6, right column).

We carried out a number of measurements with the tapered devices with the output apertures of 30 and 40 microns at different driving conditions. The results of the determination of $g^{(1)}(r)$ are shown in Figure 7.

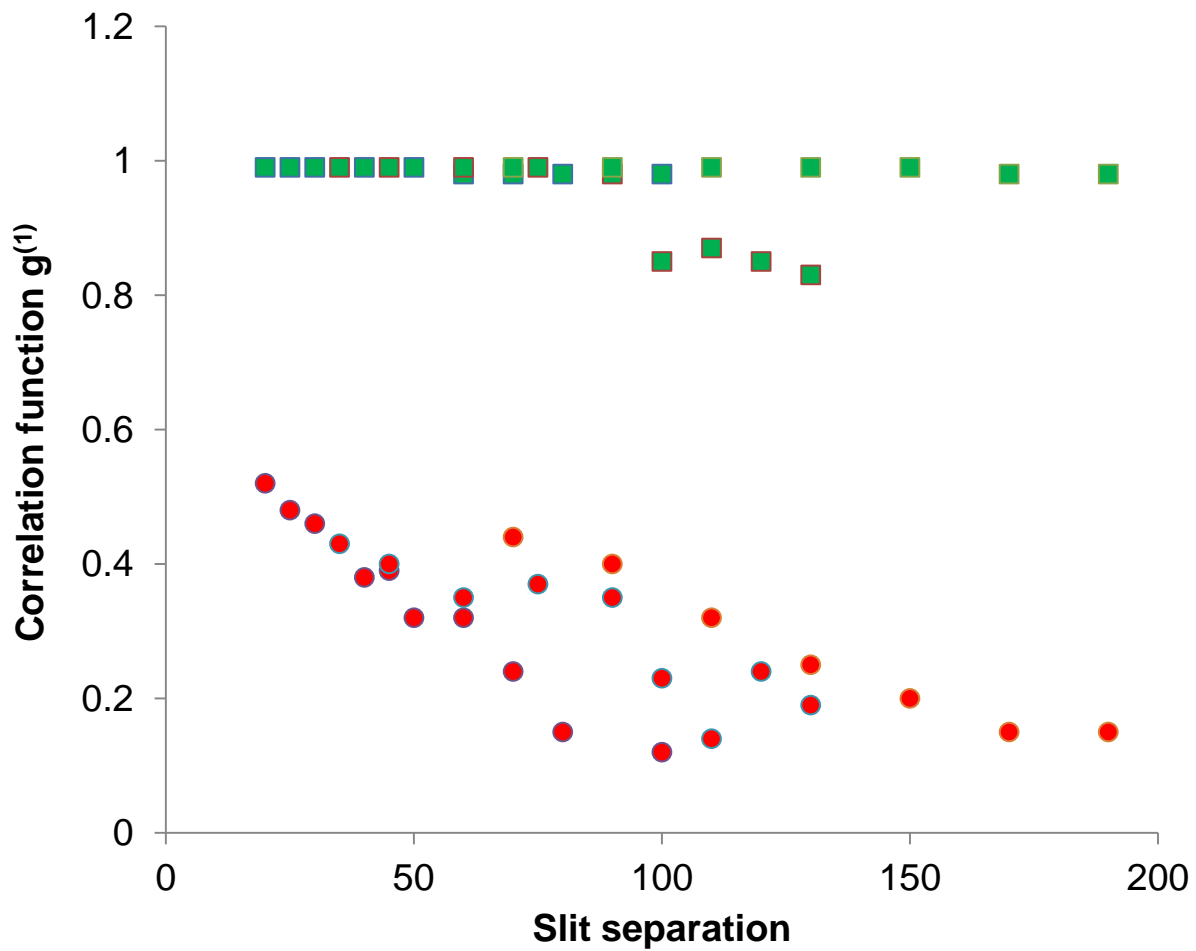

Figure 7. Correlation function $g^{(1)}$ against slit separation for lasing (red) and SR (green)

The strong difference between SR and the lasing emission is obvious. The partial spatial coherence in case of lasing can be understood by taking into account a) multiple transverse modes in the spatial profile above the laser threshold and b) temporal and spatial instabilities of the emission as shown in Fig. 4. By contrast, the first-order correlation function of the superradiant state is close to 1 at all values of the slit separation. This is an essential character of any coherent state with off-diagonal long-range order.

At this stage we should attract an attention to a very important fact. The total spatial coherence with $g^{(1)}(r) \sim 1$ can be achieved in a single (transverse) mode laser. However, the coherence is established in the laser due to the stimulated emission plus the effect of the optical feedback from the laser mirrors. If one considers the evolution of the spatial and temporal coherence, he can see how the coherent emission develops from random spontaneous noise during the travelling wave amplification when the light bounces back and forth between the mirrors. The build-up of the coherence requires hundreds and thousands of round trips within the cavity, the laser active medium being incoherent at all times. For the coherence of the SR state we have different situation. The coherence is spontaneous without any coherent external pumping and is a result of self-organization of the e-h system. Stimulated emission plays an important role in the establishing of correlations and ordering in the e-h ensemble during the early stages of the evolution of the SR state[11,17]. However, the observed interference patterns are a result of the coherence and long-range order established in the e-h system rather than the effect of the optical feedback. Typical durations of SR pulses from the tapered structures under test are below1 ps, whereas the round trip time exceeds 12 ps. This means that, by contrast to lasing, the optical feedback plays absolutely no role during the emission of the SR pulses and that the coherence of the emission originates from the coherence of the medium.

**Conclusion**

In summary, we have experimentally demonstrated the long-range temporal and spatial coherence of a macroscopically ordered state of a high-density e-h system in GaAs during the generation of femtosecond superradiant emission at room temperature. The build-up of the spatial coherence and long-range order in the dense e-h system has been observed in longitudinal and transverse directions using the Michelson interferometer and the classic Young's double slit experiment. The first-order spatial correlation function $g^{(1)}(r)$, which quantifies the off-diagonal long-range order, has been estimated of around 1 for the superradiant state. The present results give an additional evidence proving the concept of the formation of a non-equilibrium e-h BCS-like coherent state in GaAs during the superradiant emission generation developed earlier[11,16-18].

We acknowledge support by EPSRC. The authors thank V.Olle for help and H.Kan and H.Ohta for the fabrication of the semiconductor structures.

1. Anderson, M.H. *et al*, Observation of Bose-Einstein condensation in a dilute atomic vapour. *Science* **269**, 198-201 (1995).
2. Davis, K.B. *et al*, Bose-Einstein condensation in a gas of sodium atoms, *Phys. Rev. Letts*. **75**, 3969-3973 (1995).
3. Deng, H. *et al*, Condensation of semiconductor microcavity exciton polaritons, *Science* **298**, 199-202 (2002).
4. Kasprzak, J. *et al*, Bose-Einstein condensation of exciton polaritons, *Nature* **443**, 409-414 (2006).
5. Demokritov, S.O. et al, Bose-Einstein condensation of quasi-equillibrium magnons at room temperature under pumping, *Nature* **443**, 430-433 (2006).
6. Snoke, D., Coherent questions, *Nature* **443**, 403-404 (2006).
7. Ritter, S. et al, Observing the formation of long-range order during Bose-Einstein condensation, *Phys. Rev. Letts*. **98**, 090402 (2007).
8. Dicke, R.H., Coherence in spontaneous radiation processes, *Phys. Rev*. **93**, 99-110 (1954).
9. Wang, Y.K. & Hioe, F.T., Phase transition in the Dicke model of superradiance, *Phys. Rev*. A **7**, 831-836 (1973).
10. Baumann, K. et al, Dicke quantum phase transition with a superfluid gas in an optical cavity, *Nature* **464**, 1301-1306 (2006).
11. Vasil'ev, P.P., Femtosecond superradiance in inorganic semiconductors, *Rep. Prog. Phys*. **72**, 076501 (2009).
12. Gross, M. & Haroche, S., Superradiance: an essay on the theory of collective spontaneous emission, *Phys. Rep*. **93**, 301-396 (1982)
13. Andreev, A.V. et al, *Cooperative Effects in Optics: Superradiance and Phase Transitions* (IOP Publishing, Bristol, 1993).
14. Vasil'ev, P.P., Superfluorescence in semiconductor lasers, *Quantum Electron.* **27,** 860-868 (1997).
15. Vasil'ev, P.P., Role of a high gain of the medium in superradiance generation and in observation of coherent effects in semiconductor lasers, *Quantum Electron.* **29,** 842-848 (1997).
16. Vasil'ev, P.P. et al, Experimental evidence of condensation of electron-hole pairs at room temperature during femtosecond cooperative emission, *Phys. Rev B* **64**, 195209 (2001).
17. Vasil'ev, P.P., Conditions and possible mechanism of condensation of e–h pairs in bulk GaAs at room temperature, Phys. *Stat.Solidi (b)* **241**, 1251-1260 (2004).
18. Vasil'ev, P.P. & Smetanin, I.V., Condensation of electron-hole pairs in a degenerate semiconductor at room temperature, *Phys. Rev.* B **74** 125206 (2006).
19. Young, T., The Bakerian lecture: experiments and calculations relative to physical optics, *Phil. Trans. R. Soc. Lond*. **94**, 1-16 (1804).
20. Bloch, I. et al, Measurement of the spatial coherence of a trapped Bose gas at the phase transition, *Nature* **403**, 166-170 (2000).
21. Deng H. et al, Spatial coherence of a polariton condensate, *Phys. Rev. Letts*. **99**, 126403 (2007).
22. Vasil'ev, P., *Ultrafast diode lasers: fundamentals and applications* (Artech House, Norwwod, 1995).
23. Allen, L. & Eberly, J.H., *Optical resonance and two-level atoms* (Wiley, N.Y., 1975).
24. Born, M. & Wolf, E., *Principles of optics* (Cambridge University Press, Cambridge, UK, 1997).